\documentclass[reprint,twocolumn,pre,showpacs,amsmath,amssymb,aps]{revtex4-1}

\usepackage{natbib}	
\usepackage{graphicx,color,rotating}
\usepackage[latin1]{inputenc}
\usepackage{textcomp}
\usepackage{dcolumn}
\usepackage{bm}     
\usepackage{upgreek}
\usepackage{subdepth}  			
\usepackage[latin1]{inputenc}

\usepackage{array}
\newcolumntype{L}[1]{>{\raggedright\let\newline\\\arraybackslash\hspace{0pt}}m{#1}}
\newcolumntype{C}[1]{>{\centering\let\newline\\\arraybackslash\hspace{0pt}}m{#1}}
\newcolumntype{R}[1]{>{\raggedleft\let\newline\\\arraybackslash\hspace{0pt}}m{#1}}

\usepackage{hyperref}						
\usepackage{xcolor}						
\hypersetup{							
    colorlinks,
    linkcolor={blue!50!black},
    citecolor={blue!30!black},
    urlcolor={blue!80!black}
}
\newcommand{\mr}[1]{\ensuremath{\mathrm{#1}}}
\renewcommand{\vec}[1]{\bm{#1}}
\newcommand{\ee}{\mathrm{e}}
\newcommand{\ii}{\mathrm{i}}
\newcommand{\dm}{\mathrm{d}}

\newcommand{\avr}[1]{\big\langle #1 \big\rangle}

\newcommand{\pp}{\partial}

\newcommand{\nablabf}{\boldsymbol{\nabla}}

\newcommand{\etal}{\textit{et~al.}}

\newcommand{\FFF}{\vec{F}}

\newcommand{\Fgrav}{F^{\mathrm{grv}}}
\newcommand{\FFFrad}{\vec{F}^\mathrm{rad}}
\newcommand{\Frad}{F^{\mathrm{rad}}}
\newcommand{\Fradmax}{\Frad_\mathrm{max}}

\newcommand{\FFFradlat}{\FFF^\mr{rad}_\parallel}
\newcommand{\Fradlat}{F^\mr{rad}_\parallel}

\newcommand{\Hgl}{H_\mr{gl}}

\newcommand{\nnn}{\vec{n}}

\newcommand{\rrr}{\vec{r}}

\newcommand{\uuu}{\vec{u}}

\newcommand{\vvv}{\vec{v}}

\newcommand{\vflowmax}{v^\mathrm{flow}_\mathrm{max}}

\newcommand{\zerovec}{\boldsymbol{0}}

\newcommand{\Eac}{E_\mathrm{ac}}

\newcommand{\kapPS}{\kappa_\mathrm{ps}}

\newcommand{\Qflowmax}{Q^\mathrm{flow}_\mathrm{max}}

\newcommand{\etafl}{\eta_\mr{fl}}

\newcommand{\Gams}{\Gamma_\mathrm{sl}}

\newcommand{\fres}{f_\mathrm{res}}

\newcommand{\rhoPS}{\rho_\mathrm{ps}}

\newcommand{\SICel}{^\circ\!\textrm{C}}

\newcommand{\SIMHz}{\textrm{MHz}}

\newcommand{\SImuL}{\textrm{\textmu{}L}}

\newcommand{\SImm}{\textrm{mm}}
\newcommand{\SImum}{\textrm{\textmu{}m}}

\newcommand{\SIs}{\textrm{s}}

\newcommand{\beq}[1]{\begin{equation} \eqlab{#1}}
\newcommand{\eeq}{\end{equation}}
\newcommand{\bsub}{\begin{subequations}}
\newcommand{\esub}{\end{subequations}}
\def\bal#1\eal{\begin{align}#1\end{align}}
\def\bsubal#1\esubal{\bsub \begin{align}#1\end{align} \esub}

\newcommand{\eqlab}[1]{\label{eq:#1}}
\renewcommand{\eqref}[1]{Eq.~(\ref{eq:#1})}
\newcommand{\eqrefnoEq}[1]{(\ref{eq:#1})}
\newcommand{\eqsref}[2]{Eqs.~(\ref{eq:#1}) and~(\ref{eq:#2})}
\newcommand{\eqsrefnoEq}[2]{(\ref{eq:#1}) and~(\ref{eq:#2})}

\newcommand{\figref}[1]{Fig.~\ref{fig:#1}}
\newcommand{\figsref}[2]{Figs.~\ref{fig:#1} and~\ref{fig:#2}}
\newcommand{\figlab}[1]{\label{fig:#1}}

\newcommand{\secref}[1]{Section~\ref{sec:#1}}

\newcommand{\seclab}[1]{\label{sec:#1}}
\newcommand{\tabref}[1]{Table~\ref{tab:#1}}

\newcommand{\tablab}[1]{\label{tab:#1}}

\newcommand{\Eacfl}{E_\mr{ac}^\mr{fl}}
\newcommand{\Eacsl}{E_\mr{ac}^\mr{sl}}
\newcommand{\cfl}{c_\mr{fl}}

\newcommand{\gref}{g_\mr{ref}}
\newcommand{\Hg}{H_\mr{gl}}
\newcommand{\kPML}{k_\mr{PML}}
\newcommand{\Lp}{L_\mr{p}}
\newcommand{\Lpx}{L_{\mr{p},x}}
\newcommand{\Lpy}{L_{\mr{p},y}}
\newcommand{\LPML}{L_\mr{PML}}
\newcommand{\Lch}{L_\mr{ch}}
\newcommand{\Rcu}{R_\mr{cu}}
\newcommand{\dmesh}{d_\mr{mesh}}
\renewcommand{\mr}[1]{\mathrm{#1}}

\newcommand{\rhofl}{\rho_\mr{fl}}
\newcommand{\kapfl}{\kappa_\mr{fl}}
\newcommand{\rhosl}{\rho_\mr{sl}}

\newcommand{\sigmabfsl}{\boldsymbol{\sigma}_\mr{sl}}

\newcommand{\cT}{c_\mr{tr}}
\newcommand{\cL}{c_\mr{lo}}
\newcommand{\lamL}{\lambda_\mr{lo}}

\begin{document}

\title{Three-Dimensional Numerical Modeling of Acoustic Trapping in Glass Capillaries}

\author{Mikkel W. H. Ley}
\email{mley@fysik.dtu.dk}
\affiliation{Department of Physics, Technical University of Denmark, DTU Physics Building 309, DK-2800 Kongens Lyngby, Denmark}

\author{Henrik Bruus}
\email{bruus@fysik.dtu.dk}
\affiliation{Department of Physics, Technical University of Denmark, DTU Physics Building 309, DK-2800 Kongens Lyngby, Denmark}

\date{13 April 2017}

\begin{abstract}
Acoustic traps are used to capture and handle suspended microparticles and cells in microfluidic applications. A particular simple and much-used acoustic trap consists of a commercially available, millimeter-sized, liquid-filled glass capillary actuated by a piezoelectric transducer. Here, we present a three-dimensional numerical model of the acoustic pressure field in the liquid coupled to the displacement field of the glass wall, taking into account mixed standing and traveling waves as well as absorption. The model predicts resonance modes well suited for acoustic trapping, their frequencies and quality factors, the magnitude of the acoustic radiation force on a single test particle as a function of position, and the resulting acoustic retention force of the trap. We show that the model predictions are in agreement with published experimental results, and we discuss how improved and more stable acoustic trapping modes might be obtained using the model as a design tool.
\end{abstract}




\maketitle

\section{Introduction}
\seclab{Intro}
Microscale acoustofluidic devices are used increasingly in biology, environmental and forensic sciences, and clinical diagnostics \cite{Bruus2011c, Laurell2014}. Examples include cell synchronization \cite{Thevoz2010}, enrichment of prostate cancer cells in blood \cite{Augustsson2012}, manipulation of \textit{C. elegans} \cite{Ding2012}, and single-cell patterning~\cite{Collins2015}. Acoustics can also be used for non-contact microfluidic trapping and particle enrichment \cite{Hammarstrom2010, Hammarstrom2012, Hammarstrom2014}. Trapping, an important unit operation in sophisticated cell and bioparticle handling systems, can also be obtained using other technologies such as in hydrodynamic \cite{Zhang2016}, electro- and dielectro-phoretic \cite{Voldman2006}, and optical trapping systems \cite{Grier2003}, see the review by \citet{Nilsson2009}. However, one of the key parameters for successful commercialization of these technologies is throughput, and here only the hydrodynamic and acoustic systems are competitive~\cite{Cetin2014}. Hydrodynamic systems offer simple device designs and can support high flow rates, as they rely solely on passive transport from inertial migration and drag induced by secondary flows \cite{Zhang2016}. However, they do not have the capabilities of active transport.

In contrast, acoustofluidics does support active migration, basically due to the acoustic contrast between the suspending medium and the particles \cite{Karlsen2015}, and it is therefore an inherently label-free technique.  Moreover, it exhibits high cell-viability, even over several days \cite{Wiklund2012b}. These aspects and other appealing traits have increased the interest in acoustofluidic systems in recent years \cite{Bruus2011c, Laurell2014}.

{{{A sub-class of acoustofluidic devices is based on simple, cheap and commercially available glass capillaries}}}. Early studies {{{from the 1990s}}} include improving agglutination between erethrocytes \cite{Grundy1993}, simple non-contact particle manipulation in a circular wave tube by acoustic streaming and radiation forces \cite{Vainshtein1996}, and highly sensitive size-selective particle separation \cite{Wiklund2001}. However, the ultrasound capillary systems remains an active research topic, as cell-handling applications have become more numerous and refined over the years. Recent examples include cardiac myoblast viability in ultrasonic fields \cite{Ankrett2013}, sonoporation \cite{Carugo2011}, squeezing of red blood cells \cite{Mishra2014}, and trapping and retaining particles against an external flow \cite{Gralinski2014}.

\begin{figure}[b]
\centering
\includegraphics[width=0.85\columnwidth,clip]{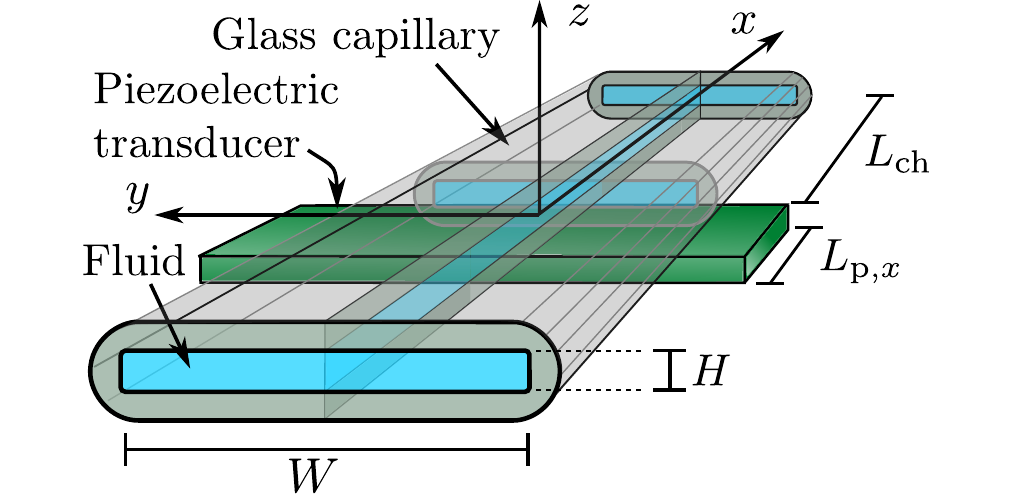}
\caption{\figlab{system_sketch}  A sketch of a generic capillary system for acoustic trapping. A mm-sized glass capillary (gray) filled with water (blue), is attached to a piezoelectric transducer (green) operated in the MHz range. At certain frequencies a resonant acoustic field builds up, in which case the acoustic radiation forces are strong enough to trap suspended microparticles in all spatial directions, and even retain them against drag from an axial fluid flow through the capillary.}
\end{figure}

In this paper, we study in particular a mm-sized glass capillary system used as a versatile acoustic trap in many experimental studies \cite{Carugo2011, Mishra2014, Lei2013, Hammarstrom2012, Hammarstrom2014} and clinical applications, such as isolation of cell-secreted membrane vesicles \cite{Evander2015} and capture and enrichment of bacteria from blood samples for rapid sepsis diagnostics \cite{Ohlsson2016}. A sketch of the generic capillary system is shown in \figref{system_sketch}, while typical material parameters are listed in \tabref{material_param}. The capillary is typically actuated locally by a piezoelectric transducer, which is coupled to the capillary either by epoxy glue or by a small drop of glycerol, the latter allowing for removal and reuse of both the capillary and the transducer. The device is driven at a resonance frequency to obtain the largest possible acoustic field in the fluid. In the optimal case, the main component of the resonance field is a vertically standing half-wave, but with an unavoidable admixture of a traveling wave in the axial direction away from the transducer. The resulting acoustic radiation forces trap particles above or near the transducer in all spatial directions and are often strong enough to retain a single microparticle against the drag from a fluid flow speed of the order of 10~$\SImum$/s.

\begin{table}[t]
\centering
\caption{\tablab{material_param} Material parameters at 25 C$^\circ$ and 4~MHz used in the numerical modeling of the capillary acoustic trap.}
\begin{ruledtabular}
\begin{tabular}{llrl}
\textit{Water} \cite{Muller2014}: &    &    & \\
Mass density     & $\rhofl$  & 997.05  & kg~m$^{-3}$ \\
Compressibility     & $\kapfl$ & 452  & TPa$^{-1}$  \\
Speed of sound  & $\cfl$ & 1496.7  & m~s$^{-1}$ \\
Dynamic viscosity & $\etafl$ & 0.890 & mPa\,s \\
Damping coefficient \cite{Hahn2015}     & $\Gamma_\mr{fl}$ & $0.004$ &  \vspace*{0.5mm} \\
Viscous boundary layer& $\delta$ & 0.27 & $\SImum$ \\
\textit{Pyrex glass} \cite{Corning_Pyrex}:& \rule{0mm}{1.1em}   &    &  \\
Mass density     & $\rhosl$  & 2230  & kg~m$^{-3}$ \\
Young's modulus     & $E$ & 62.75  & GPa  \\
Poisson's ratio  & $\nu$ & 0.2  &   \\
Speed of sound, longitudinal  & $\cL$ & 5592  & m~s$^{-1}$ \\
Speed of sound, transverse  & $\cT$ & 3424  & m~s$^{-1}$ \\
Damping coefficient  \cite{Hahn2015}    & $\Gams$ & 0.0004 &   \vspace*{0.5mm} \\
\multicolumn{4}{l}{\textit{Polystyrene} (ps) \textit{for 12~$\mu$m-diameter ps particles in water}:} \rule{0mm}{1.1em} \\
Mass density \cite{crc}    & $\rhoPS$  & 1050  & kg~m$^{-3}$ \\
Compressibility \cite{Muller2012}  & $\kapPS$ & 249  & TPa$^{-1}$  \\
Poisson's ratio \cite{Mott2008}  & $\nu_\mr{ps}$ &  0.35 &   \\
Speed of sound  at 20~$\SICel$ \cite{Bergmann1954} & $c_\mr{ps}$ & 2350  & m~s$^{-1}$ \\
Monopole coefficient, \eqref{f0} & $f_0$ & 0.48 & \\
Dipole coefficient, \eqref{f1}  & $f_1$ & 0.052-0.003i &
\end{tabular}
\end{ruledtabular}
\end{table}

In contemporary acoustofludics it is a challenge to model and optimize the design of a given device. Examples of recent advances in modeling include Lei \etal~\cite{Lei2013}, who modeled the three-dimensional (3D) fluid domain without taking the solid domain into account; Muller and Bruus \cite{Muller2014, Muller2015}, who made detailed models in 2D of the thermoviscous and transient effects in the fluid domain; Gralinski \etal~\cite{Gralinski2014}, who modeled circular capillaries in 3D with fluid and glass domains without taking absorption and outgoing waves into account; Hahn and Dual \cite{Hahn2015}, who calculated the acoustic field in a 3D model for a glass-silicon device (not a capillary system) and characterized the various loss mechanisms; and Garofalo \etal, who studied a coupled transducer-silicon-glass-water system in 2D \cite{Garofalo2017}.

In this work we present present a 3D numerical model of the capillary acoustic trap of \figref{system_sketch}. We model the acoustic pressure field in the liquid coupled to the displacement field of the glass wall, taking into account mixed standing and traveling waves as well as absorption. We model the outlets, which in practise are connected to long tubes, either as being terminated by reflecting no-stress surfaces or by absorbers that absorbs all outgoing acoustic waves from the capillary. We compare prior experimental results from the four devices listed in \tabref{geometries} with predictions of our model, in particular the frequency response, the levitating resonance modes, and the acoustic trapping forces. For one of the devices we perform two convergence analyses to show to which degree numerical convergence is obtained. Lastly, we demonstrate how to apply the model as a design tool, by studying the effects of narrowing the width of a given capillary.

\begin{table*}[]
\centering
\caption{\tablab{geometries} Capillary geometries modeled in this work: C1 from \citet{Hammarstrom2012}, C2 from \citet{Lei2013}, C3 from  \citet{Mishra2014}, C4 from \citet{Gralinski2014}, and C5 a capillary design proposed in this work. Besides the symbols defined in \figref{symm_domain}, $\Rcu$ is the radius of curvature of the fluid channel corners, $\fres^\mr{exp}$ and $\fres$ is the experimental and numerical resonance frequency, respectively, and $Q$ is the numerically calculated quality factor.}
\begin{ruledtabular}
\begin{tabular}{cccccccccccc}
 Device  &  $L$~[mm]  &  $W$~[mm] &  $H$~[mm] &  $\Hgl$~[mm] &  $\Rcu~[\SImum]$ & $\Lpx$~[mm] &  $\Lpy$~[mm]  &  $\fres^\mr{exp}$~[MHz] & $\fres$~[MHz] &   $Q$ & $d_0$~[nm]  \\
\hline  C1 & 2.0 & 2.0 & 0.20 & 0.14 & 25 & 1.16 & 2.0 & 3.970 & 3.906 &  53& 0.10\rule{0mm}{1.1em}  \\
 C2 & 2.0 & 6.0 & 0.30 & 0.30 & 25 & 1.0  & 1.0 & 2.585 &  2.495& 78& 0.10  \\
 C3 & 8.5 & 0.1 & 0.10 & 0.05 & 19 & 15   & 1.0 & 7.900 & 6.406 & 222& 0.10  \\
 C4 & 10.0 & 0.85 & 0.85 & 0.225& 425 & 4.0& 1.0 & 1.055 & 0.981 & 109& 0.10  \\
 C5 & 2.0 & 0.5 & 0.20 & 0.14 & 25 & 1.16 & 0.50& N/A   & 4.201 & 53 & 0.10
\end{tabular}
\end{ruledtabular}
\end{table*}

\begin{figure}[t]
\centering
\includegraphics[width=0.9\columnwidth,clip]{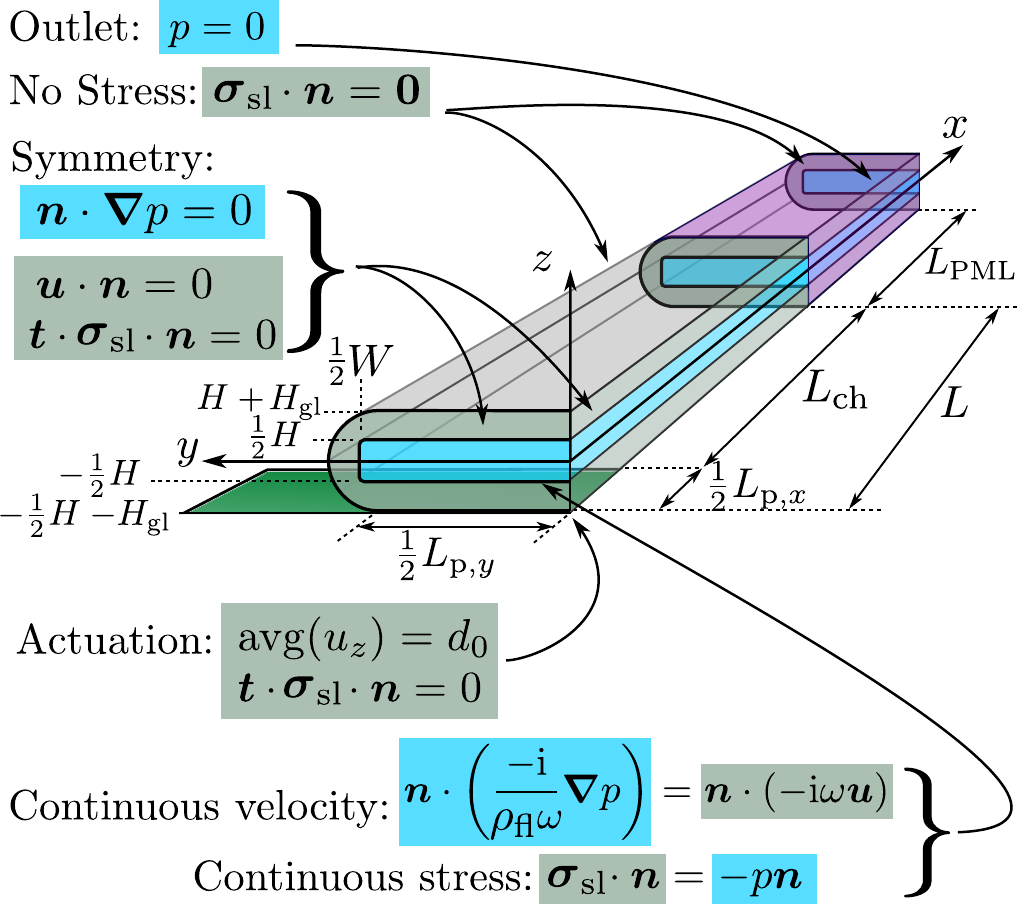}
\caption{\figlab{symm_domain}  The computational domain used in the numerical model of \figref{system_sketch} reduced to the quadrant $x > 0$ and $y > 0$ due to symmetry. Shown are the solid wall (gray), the water channel (blue), the actuation plane (green), the perfectly matched layer (PML, purple), and arrows allocating the boundary conditions to their respective boundaries.\\[-5mm]}
\end{figure}

\section{Theory and numerical model}
\seclab{theory}

We model single-frequency harmonic actuation at frequency $f$ and angular frequency $\omega = 2\pi f$,  such that any first-order acoustic field $g$ in the complex-phase representation has the time dependence $g(\rrr,t) = g(\rrr) \ee^{-\ii \omega t}$.  We introduce effective absorption in our equations by modifying the time derivative as $\pp_t \rightarrow -\ii \omega(1+\ii \Gamma)$, where $\Gamma \ll 1$ is an effective absorption parameter. The tubings at the ends of the capillary are modeled as ideal acoustic absorbers using perfectly matched layers (PML). Due to mirror symmetry about the $y$-$z$ and $x$-$z$ planes, we only model a quarter of the system as shown in \figref{symm_domain}.

\subsection{Governing equations}
The dynamics of the solid (Pyrex) is modeled by the elastic displacement $\uuu$, stress $\sigmabfsl$, density $\rhosl$, and the transverse and longitudinal speed of sound, $\cT$ and $\cL$,
 \bsub
 \bal
 \eqlab{gov_eq_solid}
 \nablabf \cdot \sigmabfsl
 &= -\rhosl \omega^2 \left(1+ \ii \Gams \right)^2 \uuu,\\
 \eqlab{solid_stress_tensor}
 \frac{1}{\rhosl}\sigmabfsl &=  \cT^2 \Big[\nablabf \uuu  + (\nablabf \uuu)^\mr{T}\Big]
 + (\cL^2 - 2 \cT^2) \big(\nablabf \cdot \uuu\big) \mr{\mathbf{I}},\\
 \eqlab{speed_of_sound_solid}
 \cT^2 &= \frac{1}{2(1+\nu)} \frac{E}{\rhosl}, \qquad
 \cL^2 = 2 \cT^2 \frac{1-\nu}{1-2\nu}.
 \eal
 \esub
Here, \textbf{I} is the unit tensor, $E$ is Young's modulus, and $\nu$ is Poisson's ratio. The fluid (water) with its acoustic pressure $p$, velocity $\vvv$, density $\rhofl$, and sound speed $\cfl$ is modeled as pressure acoustics with absorption \cite{Hahn2015},
 \bsub
 \bal
 \eqlab{gov_eq_fluid}
 \nabla^2 p &= -\frac{\omega^2}{\cfl^2} (1+\mathrm{i}\Gamma_\mr{fl})^2\: p,\\
 \vvv &= \frac{-\ii}{\omega\rhofl}\:\nablabf p.
 \eal
 \esub

\begin{table}[b]
\centering
\caption{\tablab{tab_BCs} Boundary conditions imposed on the solid and fluid domains in the model shown in \figref{symm_domain}.}
\begin{ruledtabular}
\begin{tabular}{ll}
Domain $\leftarrow$ boundary   &  Boundary condition  \\ \hline
Solid domain $\leftarrow$  air    & $\sigmabfsl \cdot \nnn = \zerovec$ \rule{0mm}{1.1em}\\
Solid domain $\leftarrow$  fluid   & $\sigmabfsl \cdot \nnn = - p\: \nnn$   \\
Solid domain $\leftarrow$  transducer   &  \eqref{rigid_act} or \eqref{avr_act} \\
Solid domain $\leftarrow$  symmetry   &   $\uuu \cdot \nnn = 0$,  $\vec{t} \cdot \sigmabfsl \cdot \nnn = 0$ \\
Fluid domain $\leftarrow$  solid  &    $\vvv \cdot \nnn  =  -\ii \omega\: \uuu\cdot\nnn$ \\
Fluid domain $\leftarrow$  air  & $p=0$ \\
Fluid domain $\leftarrow$  symmetry   & $\nnn \cdot \nablabf p = 0$ \\
\end{tabular}
\end{ruledtabular}
\end{table}

\subsection{Boundary conditions}
The applied boundary conditions are summarized in \figref{symm_domain} and \tabref{tab_BCs}. In short, the stress is zero on all outer boundaries facing the air. The stress and the velocity fields are continuous across all internal boundaries. The piezoelectric actuation on the interface $\pp\Omega_\mathrm{pz}$ is applied by imposing either an area-averaged displacement $\langle u_z\rangle_\text{pz}$ representing a transducer attached via a glycerol layer, or a rigid displacement on $u_z$ representing a transducer fixed by glue to the solid,
 \bsub
 \begin{alignat}{4}
 \eqlab{avr_act}
 \langle u_z\rangle_\text{pz} &= d_0,&\;
 \sigma_{xz} &= \sigma_{yz} &&= 0, &  \; \text{average}&\;\text{displacement},
 \\
 \eqlab{rigid_act}
 u_z &= d_0,& u_x &= u_y&& = 0, &  \text{rigid}&\;\text{displacement}.
 \end{alignat}
 \esub

Symmetry conditions are used on the two symmetry planes, while outlets with perfect absorption, which effectively removes all reflections, are modeled using perfectly matched layers as described in the following.

\subsection{Perfectly matched layers}
\seclab{PML}
In most setups, the capillary is connected to the microfluidic circuit through tubings at the ends. If a large section of the capillary is inserted into a connecting elastic tube \cite{Lei2013}, this might cause significant absorption. Furthermore, the water domain continues uninterrupted from the capillary into the connecting tube, so an extreme case  is that none of the acoustic waves going out from the actuation region are reflected at the ends of the capillary. This situation we model using perfectly matched layers (PML). A PML is a domain acting as an artificial perfect absorber of all outgoing waves, here waves traveling along the $x$ axis. In our model the region $x<L$ is normal, while the PML of length $\LPML$ is at $L<x<L+\LPML$, see \figref{symm_domain}. Following Collino and Monk \cite{Collino1998}, ideal absorption along the $x$ direction for $x>L$ can be obtained by a complex-valued coordinate stretching of $x$ in the PML domain. This stretching is based on a real-valued function $s(x)$, which is zero in the normal region $x<L$ and is smoothly increasing from zero in the PML as
 \bsub
 \eqlab{PML_definition}
 \beq{PML_absorption_func}
 s(x) = \kPML \left(\frac{x-L}{\LPML}\right)^2, \text{ for }\; L \leq x \leq L+\LPML,
 \eeq
where $\kPML>0$ is the absorption strength. The coordinate stretching is implemented in the model by changing all occurrences of $\pp_x$ and the integral measure $\dm x$ as
 \bal
 \eqlab{PML_derivative}
 \pp_x \quad \rightarrow\quad \pp_{\tilde{x}} &= \frac{1}{1+\ii s(x)}\: \pp_x,\\
 \eqlab{PML_differentials}
 \dm x \quad\rightarrow\quad \dm \tilde{x} &=  \big[1+\ii s(x)\big]\: \dm x,
 \eal
 \esub
where the latter appears in the weak-form implementation of the governing equations and boundary conditions, see \secref{ImplementationWeakForm}. The values of $\LPML$ and  $\kPML$ are also discussed in \secref{ImplementationWeakForm}. The coordinate stretching in the PML introduces a positive imaginary part in the wave numbers in the $x$ direction, and thus dampens the outgoing waves exponentially along the $x$ axis.

The opposite limit of zero absorption is handled by dropping the PML and imposing the no-stress condition at $x=L$.

\subsection{Acoustic energy density and radiation force}

Good acoustic trapping in a capillary requires a high acoustic energy density $\Eacfl$ in the fluid \cite{Pierce1991} and  $\Eacsl$ in the solid \cite{Landau1986}, both given by the sum of the space- and time-averaged kinetic and potential energy density, in the respective volumes $V_\mr{act}^\mr{fl}$ and $V_\mr{act}^\mr{sl}$ above the actuator,
 \bsubal
 \eqlab{Eac_fluid}
 \Eacfl &=  \int_{V_\mr{act}^\mr{fl}}
 \bigg[ \frac12\rhofl \avr{v^2} + \frac12\kapfl \avr{p^2}\bigg]
 \frac{\dm V}{V_\mr{act}^\mr{fl}},
 \\
 \eqlab{Eac_solid}
 \Eacsl &= \int_{V_\mr{act}^\mr{sl}} \bigg[ \frac12 \rhosl \omega^2 \avr{u^2} + \frac14
 \avr{\sigmabfsl\!:\!\big[\nablabf \uuu  + (\nablabf \uuu)^\mr{T}\big]}\bigg]
 \frac{\dm V}{V_\mr{act}^\mr{sl}},
 \esubal
where the angled bracket denotes the time average over one acoustic oscillation period.

The trapping force acting on an elastic particle of radius $a_\mr{pa}$, density $\rho_\mr{pa}$, and compressibility $\kappa_\mr{pa}$ suspended in the fluid, is the acoustic radiation force $\FFFrad$, which for mixed standing and traveling waves is given by \cite{Settnes2012}
 \bsub
 \beq{Frad}
 \FFFrad = -\frac43 \pi a_\mr{pa}^3 \Big[ \kapfl \avr{(f_0 p) \nablabf p} - \frac32 \rhofl \avr{(f_1 \vvv) \cdot \nablabf \vvv} \Big],
 \eeq
where the monopole and dipole scattering coefficients $f_0$ and $f_1$, respectively, are
 \bal
 \eqlab{f0}
 f_0 &= 1- \frac{\kappa_\mr{pa}}{\kapfl},
 \\
 \eqlab{f1}
 f_1 &= \frac{2(1-\gamma)(\frac{\rho_\mr{pa}}{\rhofl}-1)}{2\frac{\rho_\mr{pa}}{\rhofl}+1 - 3\gamma}, \ \ \gamma =-\frac{3}{2}\Big[1+\ii(1+\tilde{\delta})\Big]\tilde{\delta},
 \eal
 \esub
with $\tilde{\delta} = \frac{1}{a_\mr{pa}}\sqrt{\frac{2 \etafl}{\omega \rhofl}}$ being the viscous boundary-layer thickness normalized by the particle radius $a_\mr{pa}$.

\subsection{Numerical implementation in weak form}
\seclab{ImplementationWeakForm}
Following \citet{Gregersen2009}, we implement the governing
equations and boundary conditions in the weak-form PDE module of the finite-element software COMSOL Multiphysics 5.2 \cite{COMSOL52}. We use Lagrangian test functions of second order for $p$, $u_x$, $u_y$, and $u_z$, and solve for given geometry, materials, actuation frequency $\omega$, and actuation displacement $d_0$.

We test the PML as follows. The length $\LPML$ of the PML is comparable to the longest wavelength of the system \cite{Oskooi2008}, here the longitudinal wave length $\lamL = 2\pi \cL/\omega = 1.4$~mm in Pyrex at 4~MHz. By running parametric sweeps in $\LPML \approx \lamL$ and $\kPML > 0$, we find that for $\LPML$ as short as $0.25\lamL$, the results are independent of the PML strength in the range $10 < \kPML < 10^4$, showing that for this broad parameter range, the outgoing waves are absorbed before reaching the PML back edge at $x=L+\LPML$, and without traces of back-scattering from the PML front edge at $x=L$. To further show the independence of the results on the PML, we set $\LPML = 0.25\lamL$ and $\kPML=10^3$, and  vary the length $\Lch$ of the capillary between the piezoelectric transducer and the PML front edge. In \figref{PML_test}, $p$ and $u_z$ are plotted versus $x$ for four fixed $(y,z)$ coordinates, and it is seen that for $\Lch \gtrsim \lamL$, the resulting fields coincide everywhere outside the PML domain. For smaller values $\Lch \lesssim \lamL$ we do observe deviations in the response, indicating that the outgoing waves are not fully established before they enter the PML and are absorbed. Consequently, in the following, we choose the fixed parameter values $\Lch = 1$~mm, $\LPML = 0.25\lamL = 0.35$~mm, and $\kPML = 10^3$.

\begin{figure}[t]
\centering
\includegraphics[width=0.9\columnwidth,clip]{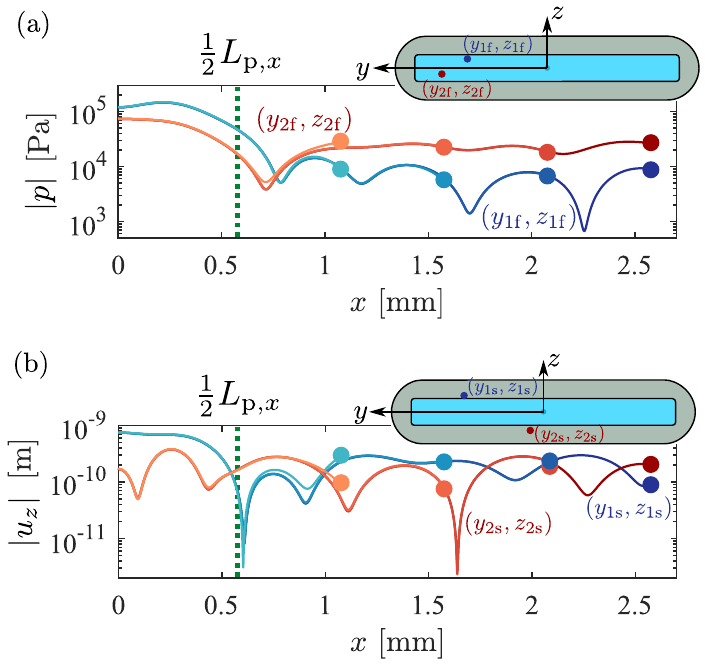}
\caption{\figlab{PML_test}  The acoustic field in capillary C1 as a function of the front-edge position $L=\frac12\Lpx+\Lch$ of the PML domain with fixed $\LPML = 0.25\lambda_\mr{L}$ and $\kPML=10^3$. (a) The pressure amplitude $|p|$ versus $x$ (colored lines) terminating at $L$ (colored points) for the two $(y,z)$ positions $(y_\mr{1f},z_\mr{1f}) = (0.3W,0.35H)$ and $(y_\mr{2f},z_\mr{2f}) = (0.4W,-0.25H)$ (blue and red points in the inset). The line colors correspond to  $\Lch = 0.5, 1.0, 1.5, 2.0$~mm from light red (light blue) to dark red (dark blue). The vertical green dotted  line at $x = \frac12\Lpx$ marks the edge of actuation region. (b) Similar plot for the vertical displacement amplitude $|u_z|$, but here for the two  $(y,z)$ positions  $(y_\mr{1s},z_\mr{1s}) = (0.3W,0.6H)$, and $(y_\mr{2s},z_\mr{2s}) = (0.05W,-0.7H)$.}
\end{figure}

The mesh is generated by first defining a mesh in the $y$-$z$ plane, see \figref{mesh}(a), and then extruding it equidistantly along the $x$ axis using the 'Swept Mesh' function in COMSOL Multiphysics, see \figref{mesh}(b). The size of the mesh elements is controlled by the maximum mesh size $\dmesh$, except elements on the $z$ axis (the intersection of the two symmetry planes) and on the corner edges of the fluid channel (where the curvature is large), both of which are assigned a smaller maximum mesh size. All mesh sizes used in the model are listed in \tabref{mesh}. Unless stated otherwise, we have set $\dmesh = H/7$.

\begin{figure}[t]
\centering
\includegraphics[width=\columnwidth,clip]{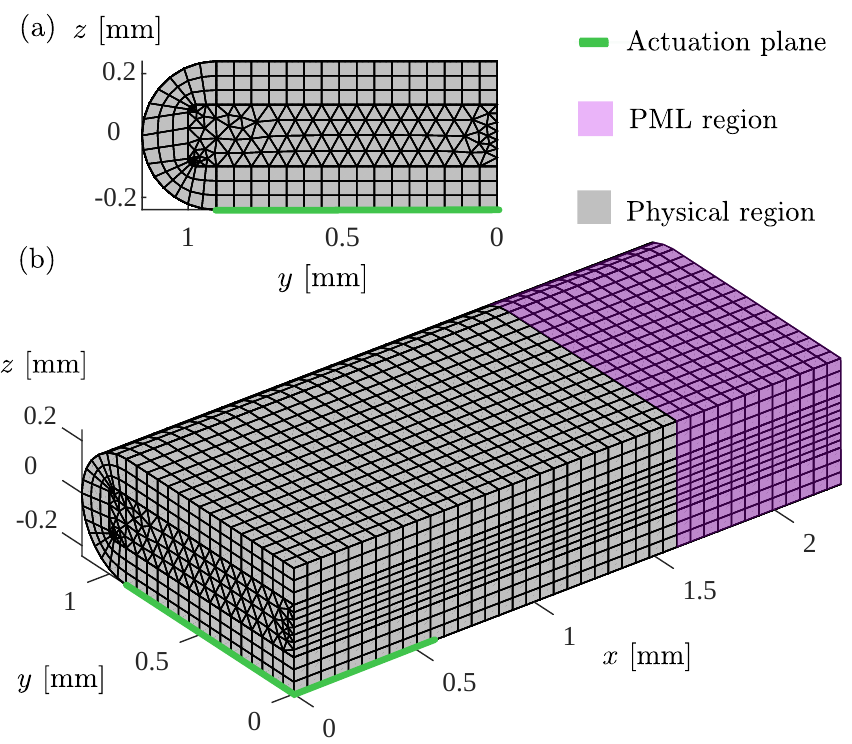}
\caption{\figlab{mesh}  A coarse version of the mesh at the surface of the computational domain of \figref{symm_domain} for a mesh size of $\dmesh = 57.2~\SImum$, which is twice the value used in most calculations, see \tabref{mesh}.  (a) The 2D mesh in the $y$-$z$ plane consisting  of a structured mesh in the solid, a free triangular mesh in the fluid,  and a refined mesh at the corners of the fluid domain as well as at the vertical symmetry line $y=0$. (b) The full 3D mesh generated by extruding the 2D mesh shown in (a) along the $x$ direction with equidistant spacing.}
\end{figure}

\begin{table}[b]
\centering
\caption{\tablab{mesh} List of the mesh parameters used in the model. The maximum mesh element size is set to $\dmesh = H/7$ with three exceptions: (1) For the reference field $g_\mathrm{ref}$ of C1 in \eqref{conv_param} and \figref{conv_test}, where $\dmesh^\mr{ref} = 0.9\dmesh$, (2) for \figref{Lorentz_line}, where $\dmesh = 40.0~\SImum$, and (3) For C4 where $\dmesh = H/20$.}
\begin{ruledtabular}
\begin{tabular}{lccc}
Location   & Linear size & No.\ of elements & Mesh size   \\
\hline
\multicolumn{4}{l}{\textit{$y$-$z$ plane}} \\
Bulk fluid  & $H$  & 7 & $\dmesh$  \\
Bulk solid    & $\Hg$  & 6 & $\dmesh$  \\
Corner solid-fluid    & $\frac12\pi \Rcu $  & 10 & 0.168\,$\dmesh$ \\
fluid symmetry edge    & $H$  & 15 & $0.5\dmesh$ \\
\multicolumn{4}{l}{\textit{$x$ direction (extruded) \rule{0mm}{1.1em}}} \\
Bulk fluid  & $L+\LPML$  & 80 & $\dmesh$  \\
Bulk solid    & $L+\LPML$  & 80 & $\dmesh$
\end{tabular}
\end{ruledtabular}
\end{table}

For a given field variable $g$, we perform a mesh convergence test based on the convergence parameter $C(g)$, which is defined in Ref.~\cite{Muller2012} as
\beq{conv_param}
C(g) = \sqrt{\frac{\int_\Omega  \dm V (g-\gref)^2}{\int_\Omega \dm V (\gref)^2}}, \text{ for}\;  0 \leq x \leq L.
\eeq
Here, the integration volume $\Omega$ is the domain in which $g$ is defined, but excluding the PML domain. The field $g$ is calculated with the above-mentioned characteristic mesh size $\dmesh$, and $\gref$ is the reference field calculated with the finer mesh $\dmesh^\mr{ref} = 0.9\dmesh$. We cannot use a smaller value of $\dmesh^\mr{ref}$, because this value combined with the length $L=2.5~\SImm$ results in a memory consumption of 85~GB RAM out of the 128~GB available on our workstation, see \secref{results}. With these values, our simulation involve $2\times 10^6$ degrees of freedom and a computation time around 15 minutes per parameter set. In \figref{conv_test}, semilog plots of $C$ versus $\dmesh^\mr{ref}/\dmesh$ for all four fields are shown. The plots exhibit an exponential decrease of $C$, which indicates good numerical mesh convergence. For the chosen mesh sizes we obtain $C \approx 0.002$, which is an acceptable level for the present study.
\begin{figure}[t]
\centering
\includegraphics[scale=1.2,clip]{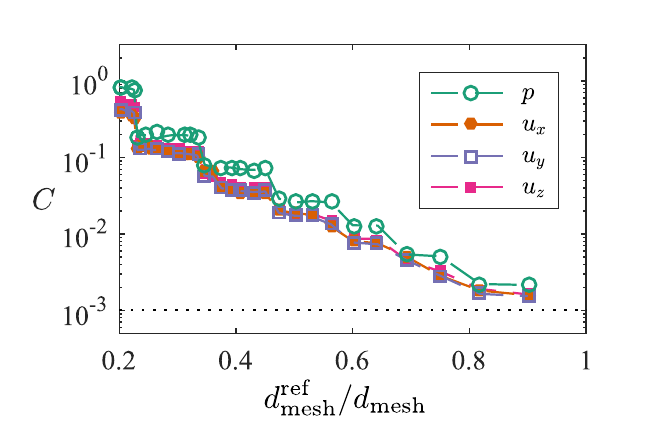}
\caption{\figlab{conv_test}  Semilog plot of the convergence parameter $C$ defined in \eqref{conv_param} versus $\dmesh^\mr{ref}/\dmesh$, the inverse of the maximum mesh size. The mesh parameter values are listed in \tabref{mesh}.}
\end{figure}

Finally, the area-averaged actuation condition \eqrefnoEq{avr_act} is imposed on the displacement field $u_z$ on the actuation interface  using an envelope function $F(x)$,
 \bsubal
 \eqlab{avg_bottom_BC}
 &\int_0^{\frac12\Lpx}\frac{2\dm x}{\Lpx}  \int_0^{\frac12\Lpy}\frac{2\dm y}{\Lpy} (u_z-d_0)\:F(x) = 0,\\
 \eqlab{env_function}
 &\text{with }\; F(x) =
 -\tanh\Bigg(\frac{x-\frac12 \Lp}{\Delta \Lp}\bigg),
 \esubal
where the transition length is set to $\Delta \Lp = 100~\SImum$.

\section{Results}
\seclab{results}

We characterize numerically the four different glass capillaries C1, C2, C3, and C4 listed in \tabref{geometries}, for which we find experimental results in the literature: C1 from \citet{Hammarstrom2012} (and also employed in Refs.~\cite{Hammarstrom2010} and \cite{Hammarstrom2014}), C2 from \citet{Lei2013}, C3 from  \citet{Mishra2014}, and C4 from \citet{Gralinski2014}. We also characterize a low-aspect ratio version of C1 denoted C5. We simulate the acoustic fields for these five capillary systems, each with the two different actuation boundary conditions \eqsrefnoEq{avr_act}{rigid_act}, and each both with and without the absorbing PML region. Supplementary to these studies, we simulate geometrical variations to study the effects of the capillaries manufacturing tolerances and the length-dependent reflections from the capillary end. This amounts to more than 20 different capillary configurations, each swept in frequency using 50 values or more, resulting in over 1000 calculations, all of which show good numerical convergence without spurious effects. We remark that a simulation of a given frequency and geometry takes between 2 and 15 minutes on our workstation, a Dell Inc Precision T3610 Intel Xeon CPU E5-1650 v2 at 3.50~GHz with 128 GB RAM and 6 CPU cores.

To gain an overall understanding of these systems, we focus in the following subsections on different acoustofluidic aspects for each capillary.

\begin{figure}[b]
\centering
\includegraphics[width=0.9\columnwidth,clip]{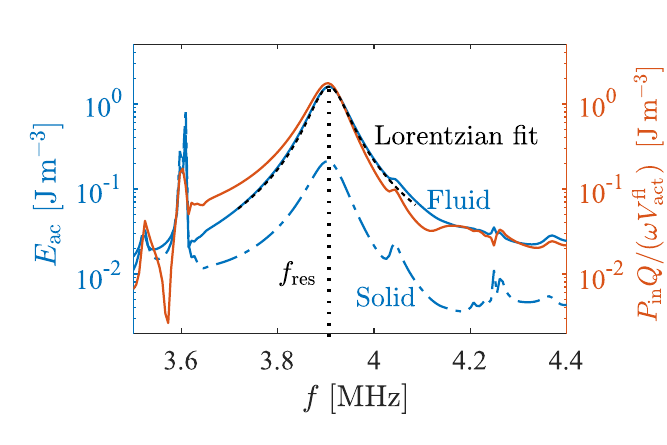}
\caption{\figlab{Lorentz_line} Semilog plot of simulation results for capillary C1 of the acoustic energy density $\Eac$ (left axis) in the fluid (blue full line) and in the solid (blue dashed line) above the actuator, and the normalized power input $Q P_\mr{in}/(\omega V^\mr{fl}_\mr{act})$ (orange line, right axis) from the actuator as a function of frequency $f$. $\Eac$ is fitted well by a Lorentzian peak (black dashed line) having a resonance frequency $\fres = 3.906~\SIMHz$, a full-width-half-maximum line width of $\Delta f =0.7369~\SIMHz$, a quality factor of $Q=53.0$, and a maximum of $\Eac = 1.60$~J/m$^3$.}
\end{figure}

\subsection{Analysis of capillary C1}
\seclab{C1}

For capillary C1 \cite{Hammarstrom2012}, we mimic the experimental use of glycerol to attach the transducer by applying the area-average condition~\eqrefnoEq{avr_act} for the actuation. Moreover, to avoid the complications arising from reflections at the end of the capillary, we first assume perfect absorption of outgoing waves and thus use the PML.

\begin{figure}[t]
\centering
\includegraphics[width=0.9\columnwidth,clip]{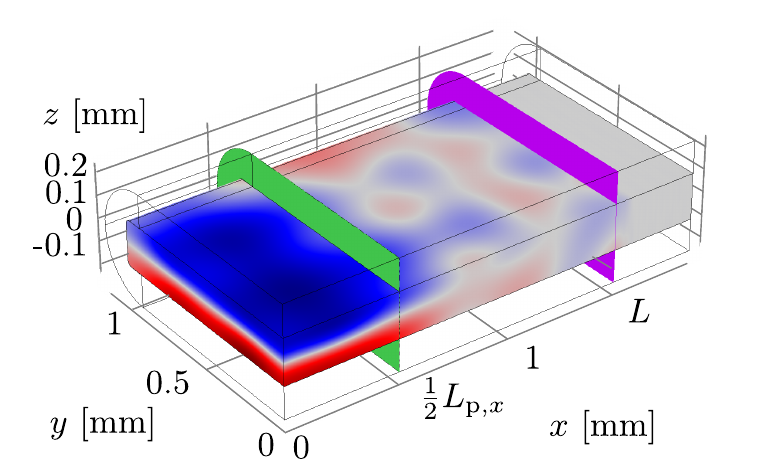}
\caption{\figlab{C1_p_field} VIDEO 1. Color plot of the pressure field $p$ in the fluid from $-0.18$~MPa (blue) to 0.18~MPa (red) of capillary C1 at the levitating half-wave resonance $\fres = 3.906~\SIMHz$. The green and purple planes represent the end of the actuation region and the beginning of the PML domain, respectively.}
\end{figure}

\begin{figure}[t]
\centering
\includegraphics[width=0.9\columnwidth,clip]{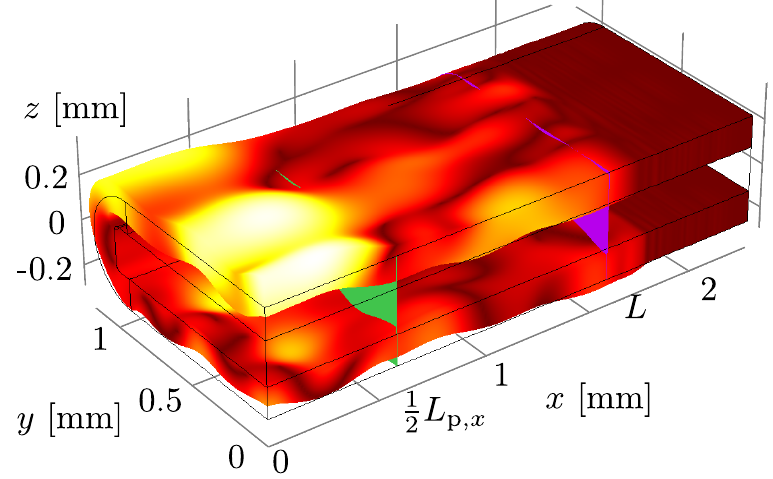}
\caption{\figlab{C1_u_field} VIDEO 2. The displacement field $\uuu$ in the solid (not drawn to scale) overlaid with a color plot of its magnitude $u$ from 0~nm (dark red) to 1.2~nm (white) for capillary C1 at the levitating resonance $\fres = 3.906~\SIMHz$ as in VIDEO 1.}
\end{figure}

First, we study the frequency response. In \figref{Lorentz_line} we show a semilog plot of the volume-averaged acoustic energy density in the fluid volume  above the actuator ($0<x<\frac12\Lpx$), $\Eacfl$ for the fluid volume $V^\mr{fl}_\mr{act}$, and $\Eacsl$ for the solid volume $V^\mr{sl}_\mr{act}$, as a function of actuation frequency $f$. A levitating resonance, strongly dominated by the energy density in the fluid, is identified at $\fres = 3.906~\SIMHz$, only 2\% lower then the experimental value $f^\mr{exp}_\mr{res}= 3.970$~MHz and 4\% from the simple half-wave value $\cfl/(2H) = 3.74$~MHz. The resonance peak is fitted well by a Lorentzian line shape centered around $\fres = 3.906~\SIMHz$ with a quality factor $Q = 53.0$.

We successfully check that this $Q$ factor is consistent with the relation $\Eac^\mr{tot} = Q P_\mr{in}/(\omega V^\mr{fl}_\mr{act})$, where $P_\mr{in} = \int_{A_\mr{act}} \avr{( -\ii \omega\uuu) \cdot \sigmabfsl\cdot \nnn} \: \dm a$ is the time-averaged power delivered by the actuator, and $\Eac^\mr{tot} = \Eacfl + \Eacsl$ is the total stored acoustic energy density. Here, we use that in steady state the input power equals the dissipated power. Besides the main levitating resonance peak, minor resonances without specific structure are also present, reflecting the many complex modes in the coupled fluid-solid system. None of these have particular good trapping properties, and they would probably not be observed in acoustophoretic experiments.

\begin{figure}[t]
\centering
\includegraphics[width=0.8\columnwidth,clip]{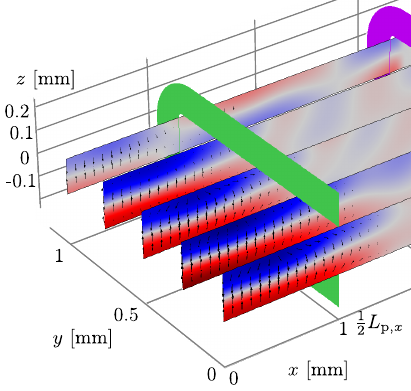}
\caption{\figlab{C1_nodal_plane}  VIDEO 3. Capillary C1, color plot of the pressure field $p$ from $-0.18$~MPa (blue) to 0.18~MPa (red) in vertical planes placed equidistantly in steps of $\frac18 W$ starting at $y=0$, showing the nearly horizontal nodal plane (gray) wobbling around $z=-0.05H$  at resonance $\fres = 3.906~\SIMHz$. The acoustic radiation force $\FFF^\mr{rad}$ on 12-$\SImum$-diameter polystyrene test particles (black arrows with logarithmic lengths for visual clarity) has a maximum magnitude of 22~pN.}
\end{figure}

Next, we study the acoustic field. Upon inspection of the pressure and displacement fields at the resonance $\fres$, we find them to be mainly a standing wave above the actuation plane combined with an outgoing traveling wave away from the actuator for $x>\frac12\Lpx$, see \figsref{C1_p_field}{C1_u_field} (VIDEO 1 and 2). The amplitude of the acoustic fields $p$ and $\uuu$ are largest above the actuator. The fluid pressure $p$ is a nearly-perfect vertical half-wave with a horizontal nodal plane near the channel center, which enables microparticle levitation given its maximum of 0.19~MPa for an actuation amplitude $d_0$ = 0.1~nm. Because of this feature, we call this type of resonance the levitating half-wave resonance in the following. At the edge of the transducer, the pressure amplitude drops an order of magnitude, which according to \eqref{Frad} gives rise to the lateral forces that constitute the acoustic trap.

For the displacement field in the solid, we find its maximum value to be 1.2~nm, which is an order of magnitude larger than the average actuation amplitude $d_0=0.1$~nm. Besides the outgoing axial displacement waves leaving the channel, we also note the presence of circumferential displacement waves with short wave lengths around $\frac14 W$. The glycerol-like, area-averaged actuation~\eqrefnoEq{avr_act} fixes only the mean vertical displacement $u_z$ while allowing for fluctuations. At resonance $\fres$ we find $u_z = (1.0\pm2.8)d_0$ on the actuation plane. We note that the standard deviation $2.8d_0 = 0.28$~nm is very small compared to a typical coupling-layer thickness of around 0.1~mm.

To characterize the trapping capabilities of the device, we study the radiation force $\FFFrad$ \eqref{Frad} acting on a 12-$\SImum$-diameter polystyrene (ps) test particle suspended in the fluid above the actuator. The $z$ component $\Frad_z$ of $\FFFrad$, derived from the nearly perfect standing half-wave in the vertical direction, levitates the test particle and holds it close to the pressure nodal plane. In the actuation region, this plane is nearly horizontal  with a vertical position between $-0.1H$ and 0. Away from the actuation region, it gradually changes its vertical position and fades away, see \figref{C1_nodal_plane} (VIDEO 3). The maximum levitating force on the test particle is $\Frad_z = 22$~pN, which is nearly 50 times larger than the buoyancy-corrected gravitational force  $\Fgrav = 0.44$~pN.

\begin{figure}[t]
\centering
\includegraphics[width=1.0\columnwidth,clip]{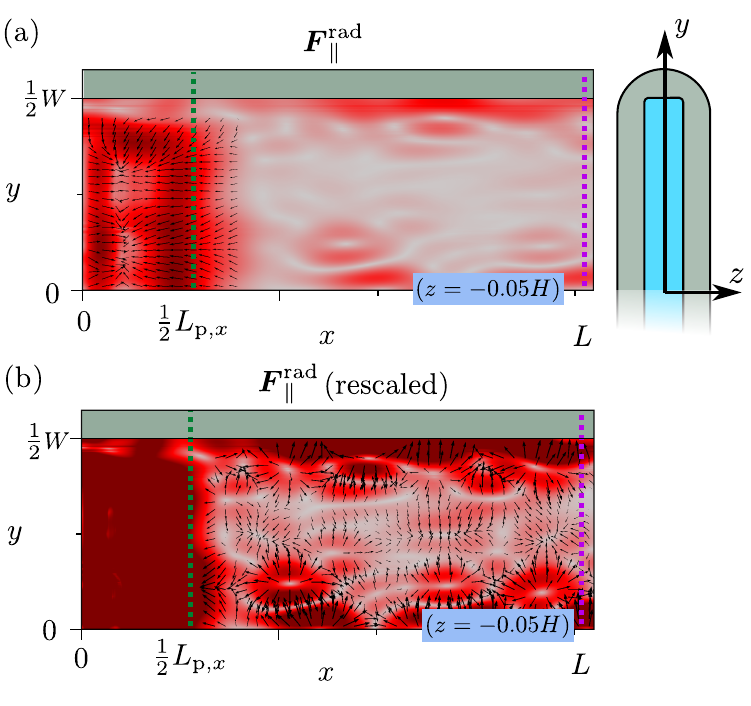}
\caption{\figlab{C1_FradPARA}  Capillary C1. (a) The lateral acoustic radiation force $\FFFradlat$ (black arrows with logarithmic length) and its magnitude (color plot from 0~pN [gray] to 0.44~pN [dark red]) acting on 12-$\SImum$-diameter polystyrene beads in the $x$-$y$ plane at $z=-0.05 H$ at resonance $\fres = 3.906$~MHz. The inset is the cross-section geometry with $W=2~\SImm$. (b) Same as (a) but with its maximum color (dark red) decreased to 0.10~pN to emphasize the secondary trapping points.}
\end{figure}

The lateral trapping force $\FFFradlat$ is given by the $x$ and $y$ components $\Frad_x$ and $\Frad_y$ of \eqref{Frad}, and it acts as a retention force when a fluid flow is imposed. In \figref{C1_FradPARA}(a) for the C1 capillary at resonance $\fres$, $\FFFradlat$ is plotted in the plane $z=-0.05H$ (vector plot) together with its magnitude $\Fradlat$ (color plot). We see two clear trapping points within the actuation region, which are surrounded by strong gradients in the $x$ and $y$ directions, thus enabling lateral particle trapping.  The maximum value $\Frad_x$ within the actuation region is of particular interest, because it serves as a measure for the overall trapping strength of the device. We find $\Frad_x = 0.44$~pN at the position $(x,y,z) = (0.3 \Lpx, 0, -0.05H)$. Outside the actuation region, a number of weaker, secondary trapping points are observed as shown in \figref{C1_FradPARA}(b). Their presence is emphasized by re-scaling the color range, and we note their positions appear to be random. This finding is in qualitative agreement with experimental observations under stop-flow conditions \cite{Ley2016b}.

We also verify the capability of the device to trap particles by simulating the trajectories of test particles initially placed on a regular grid throughout the trapping region and ending up in the trapping points (not shown).

Finally, we comment on calibration of the acoustic energy density $\Eacfl$ in the fluid, which controls the magnitude of the trapping force in the capillary device. Because in practise it is not possible to calculate $\Eacfl$ in a capillary device for a known ac voltage applied to the piezoelectric transducer, it is of importance to measure it in any given experiment. The first method is the drop voltage method \cite{Lei2013}, where at first $\Eacfl$ is set high enough to levitate a test particle. Then the ac voltage and thus $\Eacfl$ is lowered until the particle drops out of the trap, at which point $\Frad_z(\Eacfl) = \frac43\pi a^3 (\rho_\mr{pa}-\rho_\mr{fl})g$, and $\Eacfl$ can be found.

The second method is to increase the flow rate, and thus the flow speed $v_\mr{flow}$, through the device until it reaches critical magnitude, where the maximum retention force $\Fradmax$ cannot balance the Stokes drag force on the particle any longer. From a scaling argument, and introducing a coefficient $\alpha$ dependent on the device composition and geometry, we can write
 \beq{alpha_def}
 \Fradmax = \alpha \frac{4\pi}{3} a^3 \Eacfl \frac{\omega}{\cfl},\quad \alpha =  0.0139, \,
 \eeq
where the value for $\alpha$ is found by simulation on the C1 capillary. The flow speed at which the Stokes drag exactly balances $\Fradmax$ is denoted $\vflowmax$, so we can write
 \beq{force_balance}
 \Fradmax = 6\pi \etafl a \chi\vflowmax,
 \quad \chi = \chi^{z=0}_\mathrm{paral} = 1.064,
 \eeq
where $\chi$ is the wall-induced drag enhancement in the center plane of a parallel-plate channel \cite{Happel1983}, and the value is for the C1 capillary and the given test particle. Approximating the fluid channel to be rectangular, the maximum flow velocity can be related to the maximum flow rate $\Qflowmax$ as $\vflowmax = \frac{\beta}{HW}\Qflowmax$, where $\beta$ is an aspect-ratio dependent constant \cite{Bruus2008}, which for C1 is $\beta = 1.74$. Combining \eqsref{alpha_def}{force_balance} leads to
 \beq{Eac_calibration}
 \Eacfl =  \frac{9 \beta\chi}{2 \alpha }\:\frac{\etafl \cfl  \Qflowmax}{ \omega a^2 H W}.
 \eeq
With the assumed actuation amplitude $d_0 = 0.1~$nm, we find the maximum acoustic energy density to be 1.60~J/m$^3$, which yields the maximum flow speed and rate of $\vflowmax = 3.1~\SImum/\SIs$ and   $\Qflowmax = 0.043~\SImuL/$min. We emphasize that this calibration method only works if the in-plane streaming velocity is small compared to the flow velocity.

\subsection{Analysis of capillary C2}
\seclab{C2}

In the experiment, capillary C2 is held in place by two rubber-sleaves  \cite{Lei2013}, so in this case the use of the absorbing PML region is probably justified. However, capillary C2 differs from capillary C1 by being more flat, having a cross-section aspect ratio of 1:20 in contrast to that of 1:10 for the latter \cite{Hammarstrom2012}. A more significant difference is that the transducer of C2 was attached by glue and not by glycerol as for C1. We therefore study the different responses arising from using these two actuation conditions.

Using the glycerol-like area-averaged actuation condition~\eqrefnoEq{avr_act}, a strong levitating half-wave resonance resembling the one in C1 and having a quality factor $Q = 78$, is found at $f_\mr{res}^\mr{avg}=2.495~\SIMHz$ equal to the simple half-wave value $\cfl/(2H)$. The maximum levitating force is $\Frad_z = 18$~pN.

Changing to the glue-like rigid actuation condition~\eqrefnoEq{rigid_act}, the resonance frequency drops slightly to $f_\mr{res}^\mr{rgd}=2.345~\SIMHz$, fairly close to the experimentally observed resonance at 2.585~MHz, while the quality factor decreases to $Q = 61$, the acoustic energy density increases by 9\%,
and the maximum levitating force is nearly doubled to become $\Frad_z = 33$~pN. Qualitatively,  the frequency spectrum changes from having an additional resonance peak close to the levitating resonance of $\fres=2.495~\SIMHz$ using the glycerol-like condition, to having no extra resonances using the glue-like condition. The structure of the resonance field in the fluid at $\fres$ largely remains the same for the two different boundary conditions.

\subsection{Analysis of capillary C3}
\seclab{C3}

For the quadratic capillary C3  \cite{Mishra2014}, we use the nominal geometry parameters $W = H = 100~\SImum$ and $\Hgl = 50~\SImum$ with PML absorption, and we predict numerically the frequency of the levitating half-wave resonance to be $\fres = 7.35$~MHz, only 2\% from the simple half-wave value $\cfl/(2H) = 7.49$~MHz, and  7\% lower than the experimental value, $f_\mr{res}^\mr{exp} = 7.90$~MHz. This discrepancy leads us to study changes in predicted acoustic response as a function of geometrical variations within the fabrication uncertainties listed by the producer (VitroCom \#8510, Mountain Lakes, NJ): $\pm 10\%$ for $W$ and $H$, and $\pm 20\%$ for $\Hg$. Some of the resulting numerical predictions for the resonance frequency $\fres$ and $Q$ values, using both the averaged and rigid actuation condition, are listed in \tabref{fres_and_Q_C3}.

For these thin-walled capillaries, the levitating half-wave resonance depends strongly on the channel height $H$ and width $W$, and much less on the glass thickness $\Hgl$. The 10\% variation  $H=W = (100\pm10)~\SImum$ leads to a 10\% variation in $\fres \approx (7.4\pm 0.7)$~MHz for fixed $\Hgl$, while the 20\% variation in $\Hgl = (50\pm10)~\SImum$ leads to variations less than 2\% in $\fres$ for fixed $H$ and $W$. Moreover, in agreement with the results for C2, $\Eacfl$ and the $Q$ factor for the glue-like rigid actuation condition R are increased by a factor 3 - 6 and 1.2 - 3, respectively, compared to those of the glycerol-like averaged condition A. The maximum levitating force with $H=W = 100~\SImum$ for condition R is $\Frad_z = 3690$~pN, while it drops to 1380~pN for condition A. The two actuation types lead to the same levitating resonance frequency within 0.5\%.

\begin{table}[t]
\centering
\caption{\tablab{fres_and_Q_C3} Simulation results for the quadratic capillary C3:
the levitating half-wave resonance frequency $\fres$, the quality factor $Q$, and the acoustic energy density $\Eacfl$. The side length $W=H$ and glass thickness $\Hgl$ are varied within the fabrication specifications, and we use either the average actuation condition "A"~\eqref{avr_act} or the rigid condition "R"~\eqref{rigid_act}. The experimental resonance frequency is $f^\mr{exp}_\mr{res} = 7.90$~MHz.}
\begin{ruledtabular}
\begin{tabular}{cccccc}
$W = H$ & $\Hgl$ & Actu- & $\fres$ & $Q$ & $\Eacfl$ \\
$[\SImum]$ & $[\SImum]$ & ation & [MHz] & -- & [Pa] \\ \hline
 $\;\;$90 & 40 & A & 8.230 &  114 & $\;\;$67 \\
 $\;\;$90 & 60 & A & 7.964 &  $\;\;$60 & $\;\;$28 \\
 $\;\;$90 & 60 & R & 7.950 & 117 & 163 \\
100 & 50 & A & 7.350 &  \rule{2mm}{0mm}91 & $\;\;$40 \\
100 & 50 & R & 7.313 & 157 & 120 \\
110 & 40 & A & 6.670 & 124 &  $\;\;$26 \\
110 & 40 & R & 6.640 & 148 & $\;\;$84 \\
110 & 60 & A & 6.590 & \rule{2mm}{0mm}79 &  $\;\;$28 \\
\end{tabular}
\end{ruledtabular}
\end{table}

\subsection{Analysis of capillary C4}
\seclab{C4}

In the original work on the circular capillary C4 \cite{Gralinski2014}, effects due to reflections at the end of the capillary were discussed. We therefore study numerically  the frequency response with PML absorption (perfect absorption of all out-going waves) and without it (reflection of all out-going waves at the end-wall without absorption).

For the case with PML absorption, we find the levitating half-wave resonance to have the following characteristics: $\fres=  0.9753$~MHz, $Q = 109$, $\Eacfl = 0.96$~Pa, and maximum levitating force $\Frad_z = 1.6$~pN using the glycerol-like condition \eqrefnoEq{avr_act}, and $\fres=  0.9818$~MHz, $Q = 79$,  $\Eacfl = 0.57$~Pa, and maximum levitating force $\Frad_z = 2.6$~pN using the glue-like condition \eqrefnoEq{rigid_act}. These values for $\fres$ are close to the experimental value of 0.981~MHz and the theoretical value $\gamma_{11} \cfl/(\pi H) = 1.03$~MHz based on the simple hard-wall cylinder geometry, where $\gamma_{11} = 1.841$ is the first zero of the derivative $J^\prime_1(x)$ of the first Bessel function. As for the other capillaries, the choice of actuation condition leads to no qualitative, and only minor quantitative differences in the frequency spectrum.

When the PML is removed by imposing the zero-stress condition at  $x = L$, the out-going waves from the actuator region are reflected by the end wall, and complex interference effects arise in the capillary. For fixed $\Lpx$, the channel length $L$ is varied between 1~mm and 10~mm, and relative to the PML case, we find that the levitating resonance frequency only changes about 1\%, the $Q$ value generally increases and becomes $Q \approx 200 - 300$, and $\Eacfl$ increases by a factor 4 to 6.

For the longest capillary $L=10~\SImm$, we plot in \figref{C4_reflect_PML} $\Eacfl$ as a function of frequency $f$ near the levitating half-wave resonance $\fres$ with and without PML absorption. Firstly, we note that $\fres$ in the two cases differs less than 0.4\%. Secondly, without PML absorption, the energy is not lost by the out-going waves, and consequently, the line shape becomes higher and more narrow. Indeed, the resonance amplitude $\Eacfl$ increases from 0.58~Pa (PML) to 2.69~Pa (no PML), while the  $Q$ factor increases from 109 (PML) to 252 (no PML). Thirdly, without PML, the reflections from the end wall give rise to a number of additional, nearly equidistant, smaller resonance peaks, reminiscent of a Fabry-P\'erot-like interference condition for the acoustic wave in the $x$ direction. Finally, without PML, the reflections lead to a more irregular spatial pattern of the levitating resonance mode in the actuation region (not shown), as compared to that of the mode with PML absorption and no reflections.

Due to the complex nature of the coupling between pressure waves in the fluid and the compressional and shear waves in the solid, we have only been able to interpret the levitating resonance mode in simple terms of the three sound speeds of the system, listed in \tabref{material_param}, combined with the geometrical length scales of capillary C4, listed in \tabref{geometries}.

\begin{figure}[t]
\centering
\includegraphics[width=0.9\columnwidth,clip]{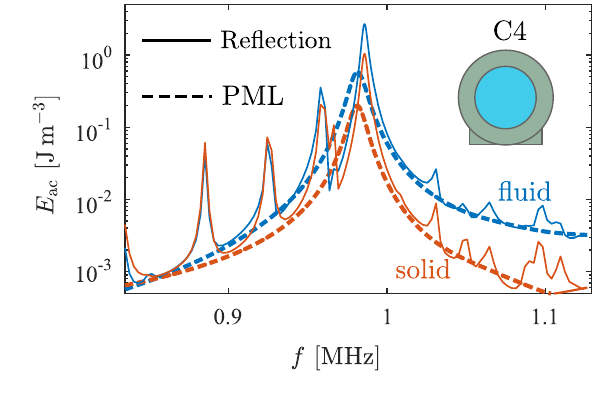}
\caption{\figlab{C4_reflect_PML} Simulated frequency response of capillary C4 (the insert shows its cross section), with total length of $L=10~\SImm$ and using the glue-like actuation condition \eqref{avr_act}. Semi-log plot of the volume-averaged acoustic energy density $\Eacfl$ in the trapping region ($x<\frac12 \Lp$) for the fluid (blue) and the solid (orange) both with PML (thick dashed lines, $\kPML=10^3$, $f_\mr{res}^\mr{PML} = 0.9812$~MHz) and without (solid lines, $\kPML=0$, $f_\mr{res}^0 = 0.9856$~MHz).}
\end{figure}

\subsection{Analysis of capillary C5}
\seclab{C5}

We end our numerical modeling of the capillaries by a study of capillary C5, which is a geometry defined by us and not studied in the literature. The focus here is to obtain a more regular spatial pattern of the mode with spatial variations on the longest possible length scale.

\begin{figure}[t]
\centering
\includegraphics[width=0.90\columnwidth,clip]{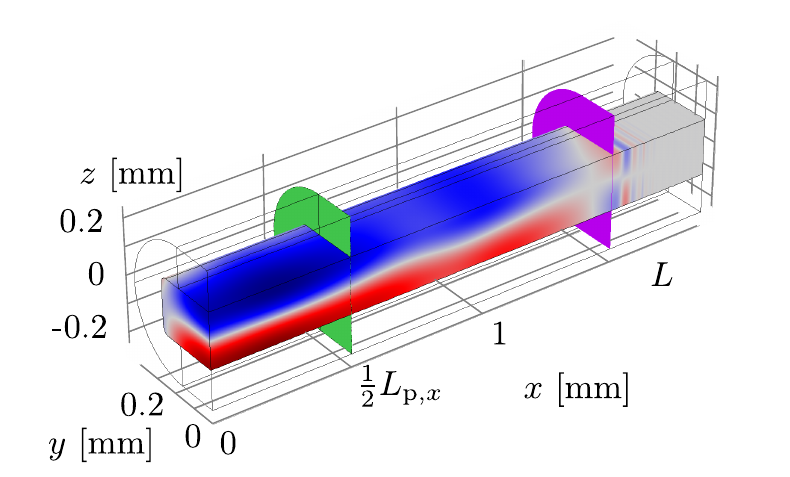}
\caption{\figlab{C5_p_field} VIDEO 4. Capillary C5, color plot of the pressure field $p$ in the fluid from $-0.10$~MPa (blue) to 0.10~MPa (red) at the resonance frequency $\fres = 4.201~\SIMHz$. The green and purple planes represent the end of the actuation region and the beginning of the PML domain, respectively.}
\end{figure}
\begin{figure}[t]
\centering
\includegraphics[width=\columnwidth,clip]{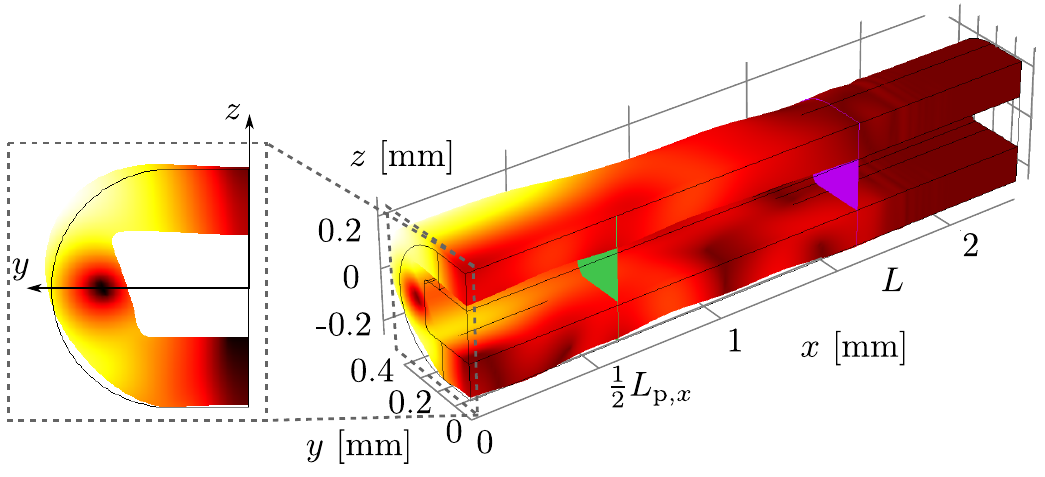}
\caption{\figlab{C5_u_field} VIDEO 5. Capillary C5, the displacement field $\uuu$ in the solid (not drawn to scale) overlaid with a color plot of its magnitude $u$ from 0~nm (dark red) to 1.2~nm (white) at resonance frequency $\fres = 4.201~\SIMHz$ as in VIDEO 3.}
\end{figure}
The study in \secref{C4} of the circular capillary C4 reveals that spatial irregularities in the levitating resonance mode are reduced when the reflections from the end wall are absorbed by a PML. This tendency we also find in the study in \secref{C1} of the wide flat capillary C1, where, even in the presence of a PML, the levitating resonance mode has significant spatial variations, see \figsref{C1_u_field}{C1_FradPARA}(b). These spatial variations are caused by the relative high mode numbers in the solid, and therefore we design capillary C5 as a version of capillary C1 with its width reduced by a factor of 4 to $W=0.5~\SImm$, thus reducing the cross-section aspect ratio $\frac{W}{H}$ from 10 to 2.5.

\begin{figure}[t]
\centering
\includegraphics[width=0.8\columnwidth,clip]{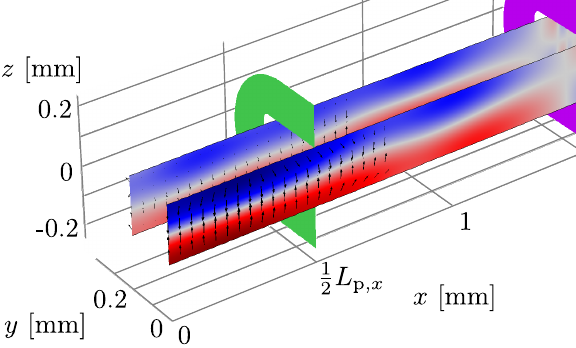}
\caption{\figlab{C5_nodal_plane} VIDEO 6. Capillary C5, color plot of the pressure field $p$ from $-0.10$~MPa (blue) to 0.10~MPa (red) in vertical planes placed equidistantly in steps of $\frac38 W$ starting at $y=0$ showing the nearly horizontal nodal plane (gray) near $z=-0.03H$ at resonance $\fres=4.201$~MHz.
The acoustic radiation force $\FFF^\mr{rad}$ on 12-$\SImum$-diameter polystyrene test particles (black arrows with logarithmic lengths for visual clarity) has a maximum magnitude of 7~pN.}
\end{figure}

\begin{figure}[t]
\centering
\includegraphics[width=\columnwidth,clip]{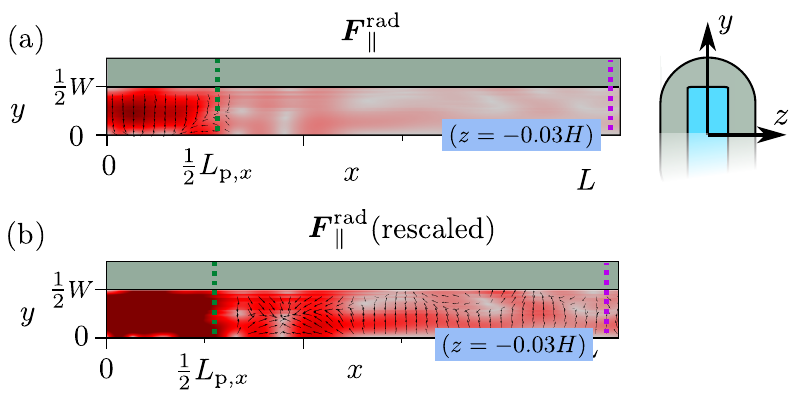}
\caption{\figlab{C5_FradPARA} Capillary C5. (a) The lateral acoustic radiation force $\FFFradlat$ (black arrows) and its magnitude (color plot from 0~pN [gray] to 0.13~pN [dark red]) acting on 12-$\SImum$-diameter polystyrene beads in the $x$-$y$ plane at $z=-0.03 H$ at resonance $\fres = 4.201$~MHz. The inset is the cross-section geometry with $W=0.5~\SImm$. (b) Same as (a) but with its maximum color (dark red) decreased to 0.05~pN (dark red) to emphasize the secondary trapping points.}
\end{figure}

The separation of resonance frequencies increases in the narrow capillary C5. In the range between 3.5 and 4.4~MHz it contains only two resonance peaks (at 3.572~MHz and 4.201~MHz), whereas capillary C1 in the same interval contains three pronounced peaks and four smaller ones. The resonance at $\fres = 4.201$~MHz, with  $Q=53$, is a levitating mode well suited for acoustic trapping, and in \figsref{C5_p_field}{C5_u_field} (VIDEO 4 and 5) we plot the corresponding pressure field $p$ and displacement field $\uuu$. The spatial structure of the resonance is more regular, especially in the solid, where the displacement in the $y$-$z$ plane now resembles a simple fundamental mode with a pivoting motion about a single stationary point away from the symmetry plane (the black point on the $y$ axis in the the inset of \figref{C5_u_field}). Compared to that of capillary C1, the levitating mode of capillary C5 stretches far outside the actuation region, it is more regular in space, but its amplitude is reduced approximately by a factor of 1.8, and the maximum levitating force drops a factor of 3 to $\Frad_z = 7$~pN. The pressure nodal surface of the levitating mode is weakly wobbling around the horizontal plane located at $z=-0.03H$, see \figref{C5_nodal_plane}. This wobbling is less pronounced than the one we find in capillary C1, which moreover is displaced farther downward, namely to $z=-0.05H$.

Compared to capillary C1, the reduced pressure amplitude in capillary C5 leads to a corresponding reduction in the lateral radiation force by a factor of $1.8^2 = 3.2$, plotted for $z=-0.03H$ in \figref{C5_FradPARA}, and it has a more regular spatial structure compared to that shown in \figref{C1_FradPARA}. The main lateral trapping points are now confined to the intersection of the nodal plane and the $x$-$z$ plane inside the actuation region, although the levitating mode extends far beyond the actuation region, see \figref{C5_nodal_plane}. There do exist secondary trapping points outside the actuation region, and compared to those of capillary C1 they are relatively strong, but they are fewer in number and appear in a more regular pattern. This finding is in qualitative agreement with experimental observations on the circular capillary C4 \cite{Gralinski2014}.

\section{Discussion}
\seclab{discussion}

The main outcome of our numerical modeling of capillary trapping devices is that for all five capillaries C1 - C5 in our study, we find a pronounced levitating mode with good trapping characteristics: a nearly horizontal pressure nodal plane near the center of the channel, see \figsref{C1_nodal_plane}{C5_nodal_plane}, with sufficiently strong lateral acoustic forces, see \figsref{C1_FradPARA}{C5_FradPARA}.

The values of the levitating resonance frequencies $\fres$ are in good agreement with the experimental values obtained for capillary C1 - C4 \cite{Hammarstrom2012, Lei2013, Mishra2014, Gralinski2014} as listed in \tabref{geometries}. They vary less than 1\% as we change the boundary condition at the outlet from an ideally absorbing PML region to the no-stress condition of a capillary facing air,  and as we change the actuation from glycerol-like area-averaged displacement to glue-like rigid displacement. However, we do find that varying these boundary conditions results in significant changes of the acoustic energy density $\Eacfl$, the acoustic force $\FFFrad$ on a test particle, as well as the resonance quality factor $Q$. In general we find two opposing trends: On the one hand the absorption of the PML region reduces both the acoustic energy and the trapping forces by a factor of 2 - 6, which is a drawback for achieving good trapping. On the other hand, the presence of the PML region has three beneficial aspects: it removes the sensitivity to reflections of acoustic waves at the outlet wall, it results in more regular spatial variations, and it makes the trapping strength independent of the position of the actuator relative to the capillary ends.

The sub-millimeter dimensions of the capillary cross section implies that the levitating resonance frequency $\fres$ is sensitive to the exact geometry. We find that when changing the geometry within the tolerances listed by the manufacturer of the capillaries, the value of the levitating resonance can vary as much as 10\% or up to 0.7~MHz for a 7-MHz resonance. Because the levitating resonance is close to an ideal standing half-wave resonance in the channel of height $H$, we have $\fres \propto H^{-1}$, and the most critical length parameter therefore becomes $H$.

Turning to the acoustic radiation force acting on a 12-$\SImum$-diameter polystyrene test particle in the device, our model correctly predicts that this force is present in the actuation region with a magnitude large enough to support one or more stable trapping points there. For the assumed actuation amplitude $d_0 = 0.1$~nm, the maximum levitating force $\Frad$ is in the range from 2~pN (the circular capillary C4) to 3690~pN (the square capillary C3), which is between 5 and 8000 times the buoyancy-corrected gravitational force $\Fgrav$, of course depending on the transducer length $\Lpx$.
Our model also correctly predicts the existence of weaker, secondary trapping points in the region outside the actuation region, see Figs.~\ref{fig:C1_FradPARA}(b) and~\ref{fig:C5_FradPARA}(b), in agreement with published experiments \cite{Gralinski2014, Ley2016b}. However, even without the PML absorption, the magnitude of the lateral acoustic force $\FFFradlat$ is small for most geometries, $\Fradlat/\Fgrav \approx 1 - 10$, the exception being capillary C3 with $\Fradlat/\Fgrav > 100 $. This indicates that the performance of actual acoustic capillary traps may depend on additional forces such as streaming-induced drag forces and particle-particle interactions.

While a wide capillary is good for enhanced throughput, it suffers from the existence of high-mode, short-wavelength elastic waves propagating around the perimeter of the capillary. We find that reducing the aspect ratio removes these perimeter waves, while maintaining the good characteristics of the levitating mode. Other irregular spatial patterns in the levitating mode and the acoustic trapping force are induced by reflections of acoustic waves at the end of the capillary. We find that by implementing an ideally absorbing PML region, these reflections are removed, and a more regular spatial behavior of the trap is obtained. While the reduction in aspect ratio reduces the through-put and the introduction of absorption reduces the trapping force by a factor 2 - 6, the resulting increased regularity in spatial behavior, the increased separation of modes in frequency implying reduced mixing of unwanted modes, and the decreased sensitivity to the exact location of the transducer relative to the capillary ends, may be beneficial design considerations worth taking into account. The re-designed devices may be more robust to small perturbations in the device geometry and mechanical actuation.

Our numerical modeling of the capillary systems for acoustic trapping is in fair agreement with existing experimental data in the literature. However, to fully establish its capability as a design tool for optimizing acoustic trapping, more experimental data is needed to characterize the acoustic properties more fully. Here, stop flow experiments as in Ref.~\cite{Ley2016b} would be helpful, as would more systematic reporting of the acoustic energy density present in the devices using, say, the drop-voltage method \cite{Lei2013} or the flow-versus-retention-force briefly mentioned in \secref{C1}. Also systematic reporting of the magnitude of the acoustic streaming would be helpful, because there might be situations in the capillary systems, where the acoustic streaming is so large that it is comparable to the flow velocities in the channel. If this were the case, then our analysis of the lateral acoustic forces must be supplemented by a discussion of the drag forces from acoustic streaming.

\section{Conclusion}

The increased use of acoustic-trapping devices in microfluidic handling of particles allows for the development of new functionalities in contemporary biotechnology. In this paper we develop a three-dimensional numerical model for water-filled glass capillaries, which may be a useful design tool in future developments of acoustic microparticle traps. We validate the model and demonstrate, how it captures many of the experimental observations reported in the literature for four different glass capillary devices, in particular regarding the frequency, the $Q$ factor, and the spatial structure and trapping capability of the levitating resonance modes.

We demonstrate the potential of the model as a design tool through an analysis of the sensitivity of the system  to changes in the geometry, the specific actuation condition, and the acoustic absorption. We find that while the existence of a main levitating resonance mode and the value of the resonance frequency are relatively insensitive to the these changes, the opposite is true for the magnitude of the trapping force and the lateral spatial structure: The trapping force is strongly increased for rectangular cross-sections with a low aspect ratio, the best being the square shape, and the lateral spatial structure of the levitating mode becomes more uniform. For the case without acoustic absorption, reflections at the outlet of the capillary make these responses very sensitive to the length of the system. The introduction of deliberate absorption of acoustic waves at the outlet would lower this sensitivity and probably result in more robust devices. However, this benefit in device operation must be weighed against a decreased magnitude of the acoustic forces.

Our numerical model takes many, but certainly not all, physical relevant aspects into account. The obvious next step, which we are currently working on, is to include the drag force from acoustic streaming as well as both acoustic and hydrodynamic particle-particle interaction effects. These more elaborate phenomena all involve lateral forces comparable to, or possibly even larger than, the lateral single-particle acoustic radiation force $\FFFrad$ calculated in our model. The simulation of these phenomena relies on numerically well-characterized acoustic fields, and this we believe is provided by the model we present here. Possible model strategies to pursue in such calculations are outlined by \citet{Muller2012, Muller2014} and  \citet{Lei2017} for the case of acoustic streaming, by \citet{Silva2014a} for acoustic particle-particle interactions, and by \citet{Ley2016} for hydrodynamic particle-particle interactions.

Our numerical model reveals that the acoustic trapping devices based on simple glass capillaries are perhaps not as simple as they look at a first glance.

\section*{Acknowledgements}
We thank our collaborators Carl Johannsson, Mikael Evander and Thomas Laurell from Lund University, Sweden, for valuable discussions and experimental data and inputs. This work was supported by Lund University and the Knut and Alice Wallenberg Foundation (Grant No. KAW 2012.0023).

%
%

\bibliographystyle{apsrev4-1-titles}

%

\end{document}